\documentclass[twocolumn,showpacs,preprintnumbers,amsmath,amssymb]{revtex4}

\usepackage{graphicx}
\usepackage{dcolumn}
\usepackage{bm}

\usepackage{amssymb}
\usepackage{amsmath}

\begin{document}


\title{Energy stored on a cosmological horizon and its thermodynamic fluctuations \\ in holographic equipartition law}

\author{Nobuyoshi {\sc Komatsu}}  \altaffiliation{E-mail: komatsu@se.kanazawa-u.ac.jp} 
\affiliation{Department of Mechanical Systems Engineering, Kanazawa University, Kakuma-machi, Kanazawa, Ishikawa 920-1192, Japan}


\begin{abstract}

Our Universe is expected to finally approach a de Sitter universe whose horizon is considered to be in thermal equilibrium.
In the present article, both the energy stored on the horizon and its thermodynamic fluctuations are examined through the holographic equipartition law.
First, it is confirmed that a flat Friedmann--Robertson--Walker universe approaches a de Sitter universe, using a cosmological model close to lambda cold dark matter ($\Lambda$CDM) models.
Then, based on the holographic equipartition law, the energy density of the Hubble volume is calculated from the energy on the Hubble horizon of a de Sitter universe.
The energy density for a de Sitter universe is constant and the order of the energy density is consistent with the order of that for the observed cosmological constant.
Second, thermodynamic fluctuations of energy on the horizon are examined, assuming stable fluctuations around thermal equilibrium states.
A standard formulation of the fluctuations for a canonical ensemble is applied to the Hubble horizon of a de Sitter universe.
The thermodynamic fluctuations of the energy are found to be a universal constant corresponding to the Planck energy, regardless of the Hubble parameter.
In contrast, the relative fluctuations of the energy can be characterized by the ratio of the one-degree-of-freedom energy to the Planck energy.
At the present time, the order of the relative fluctuations should be within the range of a discrepancy derived from a discussion of the cosmological constant problem, namely a range approximately from $10^{-60}$ to $10^{-123}$.
The present results may imply that the energy stored on the Hubble horizon is related to a kind of effective dark energy, whereas the energy that can be `maximally' stored on the horizon may behave as if it were a kind of effective vacuum-like energy in an extended holographic equipartition law.

\end{abstract}


\maketitle

\section{Introduction} 
\label{Introduction}

To explain the accelerated expansion of the late Universe \cite{PERL1998_Riess1998,Planck2015,Planck2018,Hubble2017}, various cosmological models have been proposed \cite{Bamba1}, such as lambda cold dark matter ($\Lambda$CDM) models, 
time-varying $\Lambda (t)$ cosmology, \cite{Freese-Mimoso_2015,Sola_2009-2018,Nojiri2006,Richarte2013a,Sola_2015L14,Valent2015,Sola_2017-2020,Sola2019}, bulk viscous cosmology \cite{Weinberg0,Murphy1,Barrow11-Zimdahl1,Brevik1-Nojiri1,Barrow21,Meng3,Avelino2etc2020}, 
and thermodynamic scenarios \cite{Easson,Cai,Basilakos1,Basilakos2014-Gohar,Sheykhi1,Sadjadi1,Sheykhi2,Karami2011-2016,Koma4,Koma5,Koma678,Koma9}.
The $\Lambda$CDM model assumes a cosmological constant $\Lambda$ and an additional energy component called `dark energy'.
Alternatively, thermodynamic scenarios usually assume the `holographic principle', that is, the information of the bulk is assumed to be stored on the horizon \cite{Hooft-Bousso}.
In those scenarios, the concept of black hole thermodynamics \cite{Bekenstein1,Hawking1,Hawking2} is applied to the cosmological horizon \cite{Jacob1995,Padma2010,Verlinde1,HDE,Padma2012AB,Cai2012-Tu2013,Tu2013-2015,Sheykhia2018,Neto2018a,Koma10,Koma11,Koma12,Koma14,Koma15,Koma16,Padmanabhan2004,ShuGong2011}.
The holographic equipartition law \cite{Padma2012AB} has also attracted attention and has been examined by many researchers \cite{Cai2012-Tu2013,Tu2013-2015,Sheykhia2018,Neto2018a,Koma10,Koma11,Koma12,Koma14,Koma15,Koma16}. 

In addition, the thermodynamics of the Universe has been studied from various viewpoints \cite{Jacob1995,Padma2010,Verlinde1,HDE,Padma2012AB,Cai2012-Tu2013,Tu2013-2015,Sheykhia2018,Neto2018a,Koma10,Koma11,Koma12,Padmanabhan2004,ShuGong2011,Easther1,Barrow3,Davies11_Davis0100,Gong00_01,Egan1,Pavon2013Mimoso2013,Krishna20172019,Bamba2018,Pavon2019,Saridakis2019,deSitter,Saridakis2021,Koma14,Koma15,Koma16}.
In particular, the maximization of entropy has been investigated recently \cite{Pavon2013Mimoso2013,Krishna20172019,Bamba2018,Pavon2019,Saridakis2019,deSitter,Saridakis2021,Koma14,Koma15,Koma16},
and these studies suggest that our Universe should approach a $\Lambda$-dominated universe, namely a de Sitter universe, at least in the last stage, as if our Universe behaves as an ordinary macroscopic system \cite{Pavon2013Mimoso2013}.
A de Sitter universe is considered to be in thermal equilibrium based on horizon thermodynamics.
We can expect that the energy stored on the cosmological horizon and its thermodynamic fluctuations are related to dark energy at late times, through the holographic equipartition law.
However, it is likely that the energy on the horizon and the fluctuations were not examined from this viewpoint in these works.

Thermodynamic fluctuations of the energy on the horizon should include various kinds of information.
For example, thermodynamic fluctuations on the event horizons of black holes \cite{Fluctuations1,Fluctuations2,Canonial0,OthersBlackholes} have been used in discussions on corrections of entropies \cite{Nojiri2003,Das2002,More2005,Pourhassan2017}, thermodynamic stability \cite{Stability1}, and non-Gaussian entropies \cite{AbreuNeto2020}. 
However, only a few works have studied thermodynamic fluctuations of the energy on the cosmological horizon for discussing the late Universe \cite{Mimoso2018,CosmoFluc0}. 

Of course, various fluctuations that do not occur on the horizon have been examined, such as temperature fluctuations and density fluctuations \cite{Peebles_1993,Fluc_others1,Fluc_others2,Fluc_others3}. 
Also, (quantum) vacuum fluctuations have been examined extensively \cite{CosmoFluc_Quantum,Padma2005,Elizalde2005,Verlinde2020,Ford1997}.
For example, Padmanabhan investigated vacuum fluctuations of the  energy density to explain the observed cosmological constant \cite{Padma2005}.
Verlinde and Zurek recently studied vacuum energy fluctuations on a spacetime geometry, assuming a thermal density matrix \cite{Verlinde2020}.
Vacuum energy fluctuations are characterized by an entanglement entropy discussed in anti-de Sitter/conformal field theory (AdS/CFT) \cite{Verlinde2020}, as if they are thermodynamic energy fluctuations.
These results may imply that the thermodynamic fluctuations relate to the cosmological constant problem, namely the discrepancy between the observed value of $\Lambda$ and the theoretical value of the vacuum energy estimated by quantum field theory \cite{Weinberg1989etc}. 

An understanding of both the energy stored on the cosmological horizon and its thermodynamic fluctuations should provide new insights into alternative dark energy.
In this context, we examine the energy on the cosmological horizon and its thermodynamic fluctuations, focusing on a de Sitter universe, which is considered to be in thermal equilibrium.
(The present study focuses on the late Universe and the thermodynamic fluctuations of the energy on the horizon.
The early Universe and other fluctuations are not discussed here.)

The remainder of the present article is organized as follows.
In Sec.\ \ref{Holographic models}, using a cosmological model, we study whether a flat Friedmann--Robertson--Walker (FRW) universe should approach a de Sitter universe.
In Sec.\ \ref{Thermodynamics on the horizon}, horizon thermodynamics and the holographic equipartition law of energy are reviewed.
In addition, the energy density is calculated from the energy stored on the Hubble horizon of a de Sitter universe.
In Sec.\ \ref{Fluctuations of energy}, thermodynamic fluctuations of the energy on the horizon are examined, assuming stable fluctuations around thermal equilibrium states.
In Sec.\ \ref{Standard formulations}, a standard formulation of the thermodynamic fluctuations for a canonical ensemble is reviewed.
In Sec.\ \ref{Fluctuations of energy on the Hubble horizon}, the formulation is applied to the case of the Hubble horizon of a de Sitter universe.
The thermodynamic fluctuations and the relative fluctuations are examined.
Finally, in Sec.\ \ref{Conclusions}, the conclusions of the study are presented.

\section{Holographic cosmological models with a power-law term} 
\label{Holographic models}

We consider a flat FRW universe and study the scale factor $a(t)$ at time $t$.
An expanding universe is assumed.
In this section, a general formulation of the cosmological equations is reviewed, based on previous works \cite{Koma9,Koma14,Koma16}.
Using the formulation, a holographic model that includes a power-law term is introduced, as a favored model close to $\Lambda$CDM models.
In addition, background evolutions of the universe in the present model are examined, to observe whether the flat FRW universe should approach a de Sitter universe.

To discuss a general formulation of the cosmological equations, we consider the formulation of a $\Lambda(t)$ model, similar to a time-varying $\Lambda (t)$ cosmology, because the $\Lambda(t)$ model is likely favored \cite{Koma16}.
The general Friedmann equation for the $\Lambda(t)$ model is given as 
\begin{equation}
 H(t)^2      =  \frac{ 8\pi G }{ 3 } \rho (t)    + f_{\Lambda}(t)            ,                                                 
\label{eq:General_FRW01_f_0} 
\end{equation} 
and the general acceleration equation is 
\begin{align}
  \frac{ \ddot{a}(t) }{ a(t) }      &=  -  \frac{ 4\pi G }{ 3 }  ( 1+  3w ) \rho (t)     +   f_{\Lambda}(t)                ,  
\label{eq:General_FRW02_f_0}
\end{align}
where the Hubble parameter $H(t)$ is defined as
\begin{equation}
   H(t) \equiv   \frac{ da/dt }{a(t)} =   \frac{ \dot{a}(t) } {a(t)}  , 
\label{eq:Hubble}
\end{equation}
and $w$ represents the equation of the state parameter for a generic component of matter, $w = \frac{ p(t) } { \rho(t)  c^2 }$.
Here, $G$, $c$, $\rho(t)$, and $p(t)$ are the gravitational constant, the speed of light, the mass density of cosmological fluids, and the pressure of cosmological fluids, respectively \cite{Koma9,Koma14,Koma16}.
For a matter-dominated universe, a radiation-dominated universe, and a $\Lambda$-dominated universe, $w$ is $0$, $1/3$, and $-1$, respectively.
In this section, we consider a matter-dominated universe, that is, $w =0$, although $w$ is retained for generality.
An extra driving term $f_{\Lambda}(t)$ is phenomenologically assumed.
Combining Eq.\ (\ref{eq:General_FRW01_f_0}) with Eq.\ (\ref{eq:General_FRW02_f_0}) yields \cite{Koma14}
\begin{equation}
      \dot{H} = - \frac{3}{2} (1+w) H^{2}  +  \frac{3}{2} (1+w)    f_{\Lambda}(t)     .  
\label{eq:Back_f}
\end{equation}

Using the above equation, we phenomenologically formulate a holographic model that includes a power-law term based on Padmanabhan's holographic equipartition law \cite{Koma11,Koma14,Koma15,Koma16}.
In this section, the holographic model is used as a favored model close to $\Lambda$CDM models, in order to observe typical background evolutions of a flat FRW universe.
For other models, see, e.g., Refs.\ \cite{Bamba1,Freese-Mimoso_2015,Sola_2009-2018,Nojiri2006,Richarte2013a,Sola_2015L14,Valent2015,Sola_2017-2020,Sola2019,Weinberg0,Murphy1,Barrow11-Zimdahl1,Brevik1-Nojiri1,Barrow21,Meng3,Avelino2etc2020,Easson,Cai,Basilakos1,Basilakos2014-Gohar,Sheykhi1,Sadjadi1,Sheykhi2,Karami2011-2016,Koma4,Koma5,Koma678} and the references therein.

Based on the holographic equipartition law, cosmological equations can be derived from the expansion of cosmic space due to the difference between the degrees of freedom on the surface and in the bulk \cite{Padma2012AB}.
An acceleration equation that includes a power-law term has been derived in Ref.\ \cite{Koma11}, by combining a power-law corrected entropy \cite{Das2008Radicella2010} with the holographic equipartition law.
The power-law term has been investigated in previous works \cite{Koma14,Koma15,Koma16}.
We use the following power-law term:
\begin{equation}
        f_{\Lambda}(t)   =   \Psi_{\alpha} H_{0}^{2} \left (  \frac{H}{H_{0}} \right )^{\alpha}  , 
\label{eq:fLpower}
\end{equation}
where $\alpha$ and $\Psi_{\alpha}$ are dimensionless constants whose values are real numbers \cite{Koma11}.
$H_{0}$ represents the Hubble parameter at the present time.
Also, $\alpha$ and $\Psi_{\alpha}$ are independent free parameters, and $\alpha < 2$ and $0 \leq \Psi_{\alpha} \leq 1 $ are considered.
That is, $\Psi_{\alpha}$ is a kind of density parameter for the effective dark energy.
(We discuss the condition $\alpha < 2$ later.)
Accordingly, the formulation of the present model is equivalent to that of a time-varying $\Lambda (t)$ cosmology, although the theoretical backgrounds are different.
A similar power series of $H$ for $\Lambda(t)$ models was examined in Refs.\ \cite{Valent2015,Sola2019}.
Substituting Eq.\ (\ref{eq:fLpower}) into Eq.\ (\ref{eq:Back_f}) yields 
\begin{align}
    \dot{H} &= - \frac{3}{2} (1+w)  H^{2}  +  \frac{3}{2}   (1+w)  \Psi_{\alpha} H_{0}^{2} \left (  \frac{H}{H_{0}} \right )^{\alpha}      \notag \\
               &= - \frac{3 (1+w) }{2} H^{2}  \left (  1 -   \Psi_{\alpha} \left (  \frac{H}{H_{0}} \right )^{\alpha -2} \right )        ,
\label{eq:Back_power_11}
\end{align}
the solution for which can be written as 
\begin{equation}  
    \left ( \frac{H}{H_{0}} \right )^{2-\alpha}  =   (1- \Psi_{\alpha})   \left ( \frac{a}{a_{0}} \right )^{ - \frac{3 (1+w) (2-\alpha)}{2}  }  + \Psi_{\alpha}      ,
\label{eq:Sol_HH0_power}
\end{equation}
where $a_{0}$ represents the scale factor at the present time.
The solution method is summarized in Ref.\ \cite{Koma14}.
When $f_{\Lambda}(t) = \Lambda / 3$, $\Lambda$CDM models are obtained from Eq.\ (\ref{eq:Sol_HH0_power}).
Substituting $\alpha =0$ and $w =0$ into Eq.\ (\ref{eq:Sol_HH0_power}) and replacing $\Psi_{\alpha}$ by $\Omega_{\Lambda}$ yields \cite{Koma14}
\begin{equation}
 \left (  \frac{H}{H_{0}} \right )^{2}  =   (1- \Omega_{\Lambda} )   \left ( \frac{a}{a_{0}} \right )^{ - 3}  + \Omega_{\Lambda}    ,
\label{eq:Sol_H_LCDM}
\end{equation}
where $\Omega_{\Lambda}$ is the density parameter for $\Lambda$, which is given by $\Lambda /( 3 H_{0}^{2} ) $. 
In a flat FRW universe, the density parameter for matter is given by $1- \Omega_{\Lambda}$, neglecting the influence of radiation \cite{Koma14,Koma15}.

A de Sitter universe corresponding to $\dot{H}=0$ (namely, constant $H$) can be obtained from the present model.
Equation\ (\ref{eq:Sol_HH0_power}) indicates that $(1+w)(2- \alpha) \geq 0$ satisfies $\dot{H}=0$ when $ a/a_{0} \rightarrow \infty$. 
We have already considered a matter-dominated universe ($w =0$) and, therefore, the model for $\alpha < 2$ should approach a de Sitter universe in the last stage.
(Note that $\alpha = 2$ is excluded because $\alpha < 2$ has been considered.
Also, of course, $w=-1$ satisfies $\dot{H}=0$, although $w=-1$ is not considered here.)

\begin{figure} [t] 
\begin{minipage}{0.495\textwidth}
\begin{center}
\scalebox{0.33}{\includegraphics{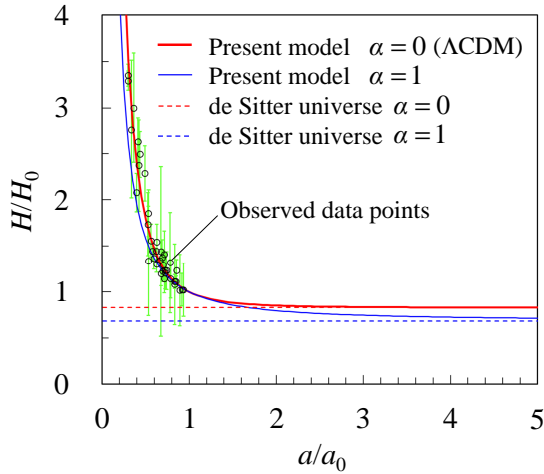}}
\end{center}
\end{minipage}
\caption{ (Color online). Evolution of the normalized Hubble parameter $H/H_{0}$ for the present model for $\alpha =0$ and $1$. 
The horizontal dashed lines represent de Sitter universes, in which a constant $H/H_{0}$ is calculated from Eq.\ (\ref{eq:Sol_HH0_deSitter}).
(See the text.)
The open circles with error bars are observed data points taken from Ref.\ \cite{Hubble2017}. 
The data points are normalized as $H/H_{0}$, where $H_{0}$ is set to $67.4$ km/s/Mpc from Ref.\ \cite{Planck2018}, as studied in Refs.\ \cite{Koma14,Koma15,Koma16}.  }
\label{Fig-H-a}
\end{figure}

We now examine typical background evolutions of the universe for the present model.
The background evolution of the universe has been examined in previous works, using a model equivalent to the present one \cite{Koma14,Koma15,Koma16}.
Based on these studies, $\alpha <2$ was found to correspond to an initially decelerating and then accelerating universe.
In particular, a model close to the $\Lambda$CDM model, namely $\alpha$ close to $0$, was favored \cite{Koma16}.
In addition, when $\alpha < 2$, the maximization of entropy was satisfied \cite{Koma16}.
Accordingly, $\alpha$ is set to $0$ and $1$, to observe typical background evolutions, as shown in Fig.\ \ref{Fig-H-a}.
Also, we set $\Psi_{\alpha}=0.685$, which is equivalent to $\Omega_{\Lambda}$ for the $\Lambda$CDM model from the Planck 2018 results \cite{Planck2018}.
Therefore, the plot for $\alpha =0$ is equivalent to that for the $\Lambda$CDM model.
In this figure, the horizontal dashed lines represent de Sitter universes, in which $H/H_{0}$ is a constant value given by 
\begin{equation}  
 \frac{H}{H_{0}}  =     \Psi_{\alpha}^{\frac{1}{2-\alpha}}      ,
\label{eq:Sol_HH0_deSitter}
\end{equation}
where we set $\alpha =0$ and $1$.
This equation is obtained by applying $ a/a_{0} \rightarrow \infty$ to Eq.\ (\ref{eq:Sol_HH0_power}) with $\alpha < 2$ and $w=0$.

As shown in Fig.\ \ref{Fig-H-a}, $H/H_{0}$ for the present model for $\alpha =0$ and $1$ decreases with $a/a_{0}$ and gradually approaches a positive value, corresponding to each de Sitter universe given by Eq.\ (\ref{eq:Sol_HH0_deSitter}).
That is, the plot for $\alpha =0$ and $1$ should approach a de Sitter universe at least in the last stage.
Accordingly, we can expect that a flat FRW universe should approach a de Sitter universe in the last stage, using a favored model close to the $\Lambda$CDM model.

In this section, a holographic model was used as one such model, in order to observe typical background evolutions.
Hereafter, we focus on a de Sitter universe.
The evolution of the normalized scale factor in a de Sitter universe is given by 
\begin{equation}  
      \frac{a}{a_{0}}  =    \exp[ H ( t - t_{0} ) ]     ,
\label{eq:aa0_deSitter}
\end{equation}
where $t_{0}$ represents the present time.
The normalized scale factor increases exponentially with time. 
Also, $H$ is not varied during the evolution of the universe.
(The properties of the de Sitter universe are characterized by $H$.)
In fact, the temperature on the cosmological horizon of a de Sitter universe is constant because the temperature depends only on $H$.
We discuss this in the next section.

\section{Horizon thermodynamics and holographic equipartition law} 
\label{Thermodynamics on the horizon}

Based on the holographic principle \cite{Hooft-Bousso}, the horizon of the universe is assumed to have an associated entropy and an approximate temperature \cite{Easson}.
In Sec.\ \ref{Bekenstein-Hawking entropy}, the entropy and the temperature on the Hubble horizon are introduced.
In Sec.\ \ref{Energy on the horizon}, the energy on the horizon is discussed, assuming the holographic equipartition law of energy.
In addition, the energy density of the Hubble volume is calculated from the energy stored on the Hubble horizon.

We focus on the thermodynamics on the horizon of a de Sitter universe.
The Hubble horizon of a de Sitter universe is equivalent to an apparent horizon because the universe is spatially flat.
For the thermodynamics of de Sitter universes, see previous works \cite{Pavon2013Mimoso2013,Krishna20172019,Bamba2018,Pavon2019,Saridakis2019,deSitter}.

\subsection{Entropy and temperature on the horizon} 
\label{Bekenstein-Hawking entropy}

Various black hole entropies have been proposed, such as power-law corrected entropy \cite{Das2008Radicella2010}, logarithmic corrections from loop quantum gravity \cite{Canonial0,MeissnerGhosh}, Tsallis--Cirto entropy \cite{Tsallis2012}, Tsallis--R\'{e}nyi entropy \cite{Czinner1,Czinner2}, and Barrow entropy \cite{Barrow2020}.
These entropies are considered to be extended versions of the Bekenstein--Hawking entropy \cite{Bekenstein1,Hawking1,Hawking2}.
Recently, Nojiri \textit{et al}. examined the thermodynamic consistency of non-Gaussian black-hole entropies \cite{Nojiri2021}.

In this study, we select the Bekenstein--Hawking entropy as an associated entropy, because it is the most standard.
In general, the cosmological horizon is examined by replacing the event horizon of a black hole by the cosmological horizon. 
The replacement has been widely accepted in previous works \cite{Jacob1995,Padma2010,Verlinde1,HDE,Padma2012AB,Cai2012-Tu2013,Tu2013-2015,Sheykhia2018,Neto2018a,Padmanabhan2004,ShuGong2011} and
we use this replacement here.

Based on a form of the Bekenstein--Hawking entropy, the entropy $S_{H}$ on the Hubble horizon is written as
\begin{equation}
 S_{H}  = \frac{ k_{B} c^3 }{  \hbar G }  \frac{A_{H}}{4}   ,
\label{eq:SBH}
\end{equation}
where $k_{B}$ and $\hbar$ are the Boltzmann constant and the reduced Planck constant, respectively. 
The reduced Planck constant is defined as $\hbar \equiv h/(2 \pi)$, where $h$ is the Planck constant \cite{Koma10,Koma11,Koma12}.
$A_{H}$ is the surface area of a sphere with the Hubble horizon (radius) $r_{H}$ given by
\begin{equation}
     r_{H} = \frac{c}{H}   .
\label{eq:rH}
\end{equation}
Substituting $A_{H}=4 \pi r_{H}^2 $ into Eq.\ (\ref{eq:SBH}) and applying Eq.\ (\ref{eq:rH}) gives
\begin{equation}
S_{H}  = \frac{ k_{B} c^3 }{  \hbar G }   \frac{A_{H}}{4}       
                  =  \left ( \frac{ \pi k_{B} c^5 }{ \hbar G } \right )  \frac{1}{H^2}  
                  =    \frac{K}{H^2}    , 
\label{eq:SBH2}      
\end{equation}
where $K$ is a positive constant given by
\begin{equation}
  K =  \frac{  \pi  k_{B}  c^5 }{ \hbar G } = \frac{  \pi  k_{B}  c^2 }{ L_{P}^{2} } = \frac{  \pi  k_{B}  }{ t_{P}^{2} }  , 
\label{eq:K-def}
\end{equation}
and $L_{P}$ is the Planck length and $t_{P}$ given by $L_{P}/c$ is the Planck time, written as
\begin{equation}
  L_{P} = \sqrt{ \frac{\hbar G} { c^{3} } }    \quad \textrm{and} \quad   t_{P} = \sqrt{ \frac{\hbar G } { c^{5} } }  .
\label{eq:Lp-tp}
\end{equation}
In addition, the Planck mass $M_{P}$, the Planck energy $E_{P} = M_{P} c^{2} $, and the Planck temperature $T_{P} = E_{P} /k_{B} $ are given as 
\begin{equation}  
M_{P} = \sqrt{ \frac{\hbar c } { G } } ,  \quad    E_{P} = \sqrt{ \frac{\hbar c^{5}} { G } }  ,  \quad   T_{P} = \frac{1}{k_{B}}\sqrt{ \frac{\hbar c^{5}} { G } }  .
\label{eq:Mp-Ep}
\end{equation}
Next, we introduce the temperature $ T_{H}$ on the Hubble horizon.
The temperature can be given by \cite{GibbonsHawking1977} 
\begin{equation}
 T_{H} = \frac{ \hbar H}{   2 \pi  k_{B}  }   .
\label{eq:T_H1}
\end{equation}
From Eqs.\ (\ref{eq:SBH2}) and (\ref{eq:T_H1}), $S_{H}$ and $T_{H}$ are proportional to $H^{-2}$ and $H$, respectively, and depend only on $H$.
Therefore, $S_{H}$ and $T_{H}$ are constant during the evolution of the universe because a de Sitter universe is considered.

It should be noted that the second law of thermodynamics and the maximization of entropy are not discussed in the present paper,
but have been examined in previous works, see, e.g., Ref.\ \cite{Koma14}.

\subsection{Holographic equipartition law of energy}
\label{Energy on the horizon}

We have assumed that the information of the bulk is stored on the horizon based on the holographic principle \cite{Hooft-Bousso}.  
In addition, we have considered a de Sitter universe, in which $T_{H}$ given by Eq.\ (\ref{eq:T_H1}) is constant.
Accordingly, the concept of a canonical ensemble can be applied to the de Sitter universe.
We now assume the equipartition law of energy on the horizon, according to Refs.\ \cite{Padma2010,ShuGong2011}.
Consequently, an average energy on the Hubble horizon, $E_{H} = \left < \mathcal{E}_{H} \right >$, can be written as
\begin{equation}
E_{H} = \left < \mathcal{E}_{H} \right > =  N_{H}  \times \frac{1}{2} k_{B} T_{H}     , 
\label{E_equip}
\end{equation}
where $ \mathcal{E}_{H} $ represents the energy on the horizon and $\left < \mathcal{E}_{H} \right >$ represents the (ensemble) average energy.
In the present study, the symbol `$\left < \quad \right >$' is used for average values, and fluctuations will be discussed later.
Note that $E_{H}$ is used as well and we call $E_{H}$ the Hubble energy for simplicity.
Also, $N_{H}$ is the number of degrees-of-freedom (DOF) on a spherical surface of the Hubble radius $r_{H}$, which is written as \cite{Koma14}
\begin{equation}
  N_{H} = \frac{4 S_{H} }{k_{B}}       .
\label{N_sur}
\end{equation}
Substituting Eq.\ (\ref{N_sur}) into Eq.\ (\ref{E_equip}) yields
\begin{equation}
\left < \mathcal{E}_{H} \right > =  \left ( \frac{4 S_{H} }{k_{B}} \right )     \frac{1}{2} k_{B} T_{H}  =  2 S_{H}  T_{H}  . 
\label{E_ST}
\end{equation}
This thermodynamic relation, namely $ \left < \mathcal{E}_{H} \right >  =2 S_{H}  T_{H} $, was proposed by Padmanabhan \cite{Padmanabhan2004,Padma2010}.
In fact, the factor $1/2$ in Eq.\ (\ref{E_equip}) should depend on the Hamiltonian of the system.
In Ref.\ \cite{Padma2010}, the factor was assumed to be $1/2$
and we have similarly assumed a factor $1/2$, which satisfies a thermodynamic relation discussed later.

We reformulate the average energy given by Eq.\ (\ref{E_ST}), using the Hubble parameter, as examined in Ref.\ \cite{Koma14}.
Substituting Eqs.\ (\ref{eq:SBH2}) and (\ref{eq:T_H1}) into Eq.\ (\ref{E_ST}) yields
\begin{equation}
\left < \mathcal{E}_{H} \right >  =  2  \left ( \frac{ \pi k_{B} c^5 }{ \hbar G } \right )  \frac{1}{H^2}  \times \left ( \frac{ \hbar H}{   2 \pi  k_{B}  }    \right )  =    \frac{ c^{5} }{ G }  \left ( \frac{1}{H} \right )  .
\label{E_ST2}
\end{equation}
The average energy $\left < \mathcal{E}_{H} \right > $ depends on $H$, which is constant in a de Sitter universe.

From the above equation, we can confirm the thermodynamic relation $\frac{ d \left < \mathcal{E}_{H} \right > }{T_{H}} = d S_{H}$.
Using $ d \left < \mathcal{E}_{H} \right > = \frac{ c^{5} }{ G }  \left (- \frac{dH}{H^{2}} \right ) $ obtained from Eq.\ (\ref{E_ST2}) and dividing $ d \left < \mathcal{E}_{H} \right >$ by Eq.\ (\ref{eq:T_H1}) yields 
\begin{equation}
\frac{ d \left < \mathcal{E}_{H} \right > }{T_{H}}  =   \frac{\frac{ c^{5} }{ G }  \left (- \frac{dH}{H^{2}} \right )  }{\frac{ \hbar H}{   2 \pi  k_{B}  }} = \left ( \frac{ \pi k_{B} c^5 }{ \hbar G } \right ) \left (- \frac{2 dH }{H^{3}} \right ) = d S_{H}  .
\end{equation}
Here $ d S_{H} = \left ( \frac{ \pi k_{B} c^5 }{ \hbar G } \right ) \left (- \frac{2 dH }{H^{3}} \right ) $ is obtained from Eq.\ (\ref{eq:SBH2}).
This thermodynamic relation has been discussed by Padmanabhan \cite{Padmanabhan2004} and Shu and Gong \cite{ShuGong2011}.

We now calculate the energy density of the Hubble volume $V_{H}$ from the average energy $ \left < \mathcal{E}_{H} \right > $ on the Hubble horizon, by applying the holographic principle.
The Hubble volume considered here is constant because the Hubble radius $r_{H}$ is constant in a de Sitter universe.
Dividing Eq.\ (\ref{E_ST2}) by $V_{H}$ and applying $r_{H} = c/H$ given by Eq.\ (\ref{eq:rH}) gives the energy density $\rho_{H} c^{2}$, written as
\begin{equation}
 \rho_{H} c^{2} = \frac{ \left < \mathcal{E}_{H} \right >   }{V_{H}}= \frac{ \frac{ c^{5} }{ G }  \left ( \frac{1}{H} \right )    }{\frac{4}{3} \pi r_{H}^3}
= \frac{ \frac{ c^{5} }{ G }  \left ( \frac{1}{H} \right )    }{\frac{4}{3} \pi (c/H)^3} =    \frac{ 3 c^{2} }{ 4 \pi G }  H^{2}   , 
\label{energy-density}
\end{equation}
or equivalently, the mass density $\rho_{H}$ of the Hubble volume is given by 
\begin{equation}
 \rho_{H}  =    \frac{ 3 }{ 4 \pi G }  H^{2}   .
\label{mass-density}
\end{equation}
The obtained $\rho_{H}$ is constant during the evolution of a de Sitter universe.
Also, the formulation of Eq.\ (\ref{mass-density}) is similar to the standard Friedmann equation, $ H^2 =  \frac{ 8\pi G }{ 3 } \rho $. 
Using the Friedmann equation, the critical mass density $ \rho_{c}$ is given by $ \rho_{c} = \frac{ 3 H^{2} }{ 8 \pi G  }$.
At the present time, $ \rho_{c}$ is expected to be slightly larger than the observed mass density $\rho_{\Lambda,obs}$ for $\Lambda$, because $\Omega_{\Lambda} = \rho_{\Lambda,obs} / \rho_{c} =0.685$ from the Planck 2018 results \cite{Planck2018}.
Therefore, the order of $\rho_{H}$ is consistent with the order of $\rho_{\Lambda,obs}$.
The relation can be written as
\begin{equation}
 \rho_{H}  =  2 \rho_{c} \approx \rho_{\Lambda,obs}  .
\label{rhoH_rhoLobs}
\end{equation}
These results may imply that the energy stored on the horizon can be considered to be the energy in the volume and can be interpreted as a kind of effective dark energy.

Again, $\rho_{H}$ given by Eq.\ (\ref{mass-density}) is constant for a de Sitter universe
and, therefore, we cannot apply $\rho_{H} =\frac{ 3 H^{2} }{ 4 \pi G }$ directly to the extra driving term $f_{\Lambda}(t)$ included in Eq.\ (\ref{eq:Back_f}).
In fact, for $f_{\Lambda}(t) \propto H^{2}$, cosmological models cannot describe a decelerating and accelerating universe \cite{Basilakos1,Koma5}.
For details on the various driving terms, see, e.g., the works of Sol\`{a} \textit{et al}. \cite{Sola_2015L14}, Valent \textit{et al}. \cite{Valent2015}, and Rezaei \textit{et al}. \cite{Sola2019}.

\section{Thermodynamic fluctuations}
\label{Fluctuations of energy}

We consider a canonical ensemble in statistical physics, assuming stable fluctuations around thermal equilibrium states \cite{Das2002,Nojiri2003,More2005,Pourhassan2017}.
In Sec.\ \ref{Standard formulations}, a standard formulation of the thermodynamic fluctuations of energy is reviewed.
In Sec.\ \ref{Fluctuations of energy on the Hubble horizon}, the formulation is applied to the case of the Hubble horizon of a de Sitter universe to examine the thermodynamic fluctuations and the relative fluctuations.

Thermodynamic stability and negative specific heat are not discussed in this paper.
Also, fluctuation theorems \cite{Evans1993etc} are not discussed here because they assume nonequilibrium steady states.

\subsection{Formulations of thermodynamic fluctuations}
\label{Standard formulations}

In this subsection, thermodynamic fluctuations of energy for general systems are introduced.
To this end, we review a standard formulation of the fluctuations, according to the work of Das \textit{et al}. \cite{Das2002}.

We consider a canonical ensemble with a partition function given by 
\begin{equation}
Z (\beta) =  \int_{0}^{\infty} \varrho (\mathcal{E}) \exp (- \beta \mathcal{E}) d\mathcal{E}    ,
\label{eq:Z}
\end{equation}
where $\varrho$ is the density of states, $\mathcal{E} $ is the energy of the system, and $\beta$ is the inverse temperature, written as
\begin{equation}
\beta  = \frac{1}{k_{B} T} , 
\label{eq_beta}
\end{equation}
where $T$ represents the temperature of the system.
Based on statistical physics, entropy $S$ is written as \cite{Das2002}
\begin{equation}
 S (\beta)   = k_{B} \ln Z(\beta) + k_{B}  \beta \mathcal{E} , 
\label{eq_Sbeta}
\end{equation}
and $S^{\prime \prime} (\beta) \equiv \frac{\partial^{2} S}{ \partial \beta^{2}}$ can be written as  
\begin{align}
 S^{\prime \prime} (\beta)
&\equiv    \frac{\partial^{2} S}{ \partial \beta^{2}}   
&= k_{B}  \left ( \frac{1}{Z} \frac{ \partial^{2} Z}{ \partial \beta^{2} } -  \left ( \frac{-1}{Z} \frac{ \partial Z}{ \partial \beta } \right )^{2}  \right )    ,
\label{S_prime2_0}
\end{align}
where $^{\prime}$ represents $\partial /\partial  \beta$.
For details, see Ref.\ \cite{Das2002}.

In addition, using statistical physics relations, the (ensemble) average of $\mathcal{E}$ in an equilibrium state at $\beta= \beta_{0}$ can be given by \cite{Das2002}
\begin{equation}
\left < \mathcal{E} \right > =   - \left [ \frac{ \partial }{ \partial \beta } \ln Z  \right ]_{\beta= \beta_{0}}  =   -  \left [  \frac{1}{Z} \frac{ \partial Z}{ \partial \beta } \right ]_{\beta= \beta_{0}} , 
\label{E_Z_ave}
\end{equation}
and the average of $\mathcal{E}^{2}$ is written as
\begin{equation}
\left < \mathcal{E}^{2} \right >  =   \left [   \frac{1}{Z} \frac{ \partial^{2} Z}{ \partial \beta^{2} }  \right ]_{\beta= \beta_{0}} ,
\label{E2_Z_ave}
\end{equation}
where $\beta_{0} = 1/(k_{B} T_{0}) $ represents the inverse temperature in the equilibrium state and $T_{0}$ is an equilibrium temperature.
Hereafter the subscript $\beta= \beta_{0}$ is omitted.

Substituting Eqs.\ (\ref{E_Z_ave}) and (\ref{E2_Z_ave}) into Eq.\ (\ref{S_prime2_0}) gives $ S^{\prime \prime} (\beta)$ at the equilibrium state \cite{Das2002}:
\begin{align}
 S^{\prime \prime} (\beta)   &= k_{B}  \left ( \frac{1}{Z} \frac{ \partial^{2} Z}{ \partial \beta^{2} } -  \left ( \frac{-1}{Z} \frac{ \partial Z}{ \partial \beta } \right )^{2}  \right ) \notag \\
                                         &= k_{B}  \left ( \left < \mathcal{E}^{2} \right >  -  \left < \mathcal{E} \right >^{2}   \right )  =  k_{B} \left <  \left ( \delta \mathcal{E} \right )^{2} \right >   , 
\label{S_prime2}
\end{align}
where a fluctuation $ \delta \mathcal{E} $ is defined by 
\begin{equation}
\delta \mathcal{E} \equiv   \mathcal{E}  - \left < \mathcal{E} \right >   ,
\label{dE_EEave}
\end{equation}
and the average of $ \delta \mathcal{E} $ is $0$, namely $< \delta \mathcal{E} > =0$.
From Eq.\ (\ref{S_prime2}), the variance $\sigma_{\mathcal{E}}^{2}$ of the energy is given by 
\begin{equation}
 \sigma_{\mathcal{E}}^{2} \equiv \left <  \left ( \delta \mathcal{E} \right )^{2} \right >  =\frac{ S^{\prime \prime} (\beta) }{ k_{B} } .
\label{variance}
\end{equation}
Using this equation, the variance $\sigma_{\mathcal{E}}^{2}$ is calculated from the entropy in the equilibrium state at $\beta= \beta_{0}$.

In this subsection, we have introduced thermodynamic fluctuations of energy for general systems.
In the next subsection, we apply Eq.\ (\ref{variance}) to the case of the fluctuations on a cosmological horizon.
(In Refs.\ \cite{Das2002,Nojiri2003,More2005,Pourhassan2017}, Eq.\ (\ref{variance}) was used for corrections of black hole entropies.)

\subsection{Thermodynamic fluctuations of the energy on the Hubble horizon of a de Sitter universe}
\label{Fluctuations of energy on the Hubble horizon}

We consider a de Sitter universe in which the temperature $T_{H}$ on the Hubble horizon is constant.
Accordingly, the Hubble horizon should be in an equilibrium state, namely $T_{H}=T_{0}$.
In this case, thermodynamic fluctuations of energy on the horizon can be calculated from Eq.\ (\ref{variance}), assuming stable fluctuations around thermal equilibrium states.

To apply Eq.\ (\ref{variance}), the entropy $S_{H}$ on the Hubble horizon is formulated as a function of $\beta$.
From Eq.\ (\ref{eq_beta}), $\beta$ is given by
\begin{equation}
\beta  = \frac{1}{k_{B} T} = \frac{1}{k_{B} T_{H}}  ,
\label{eq_beta_H}
\end{equation}
where $T=T_{0}=T_{H}$ is assumed.
Using Eq.\ (\ref{eq:SBH2}), the entropy on the Hubble horizon is written as
\begin{equation}
S_{H}   =  \left ( \frac{ \pi k_{B} c^5 }{ \hbar G } \right )  \frac{1}{H^2}    .
\label{eq:SBH3}      
\end{equation}
Also, $H$ can be calculated from $T_{H} = \hbar H /  (2 \pi  k_{B})$ given by Eq.\ (\ref{eq:T_H1}).
Solving $T_{H} = \hbar H /  (2 \pi  k_{B})$ with respect to $1/H$ and substituting $1/H =  \hbar /(2 \pi  k_{B}T_{H}) $ into Eq.\ (\ref{eq:SBH3}) yields
\begin{align}
S_{H}  &=  \left ( \frac{ \pi k_{B} c^5 }{ \hbar G } \right ) \left ( \frac{\hbar}{ 2 \pi  k_{B}T_{H} } \right )^{2}  =   \frac{ \hbar k_{B} c^5 }{ 4 \pi G }  \left ( \frac{1}{ k_{B}T_{H} } \right )^{2}    \notag \\
                  &=   \left ( \frac{ \hbar k_{B} c^5 }{ 4 \pi G } \right )   \beta^{2}    .
\label{eq:SBH4}      
\end{align}

We now examine the variance $\sigma_{\mathcal{E}_{H}}^{2}$ of the energy on the Hubble horizon.
Substituting Eq.\ (\ref{eq:SBH4}) into Eq.\ (\ref{variance}) yields
\begin{equation}
\sigma_{\mathcal{E}_{H}}^{2} 
 =\frac{ S_{H}^{\prime \prime} (\beta) }{ k_{B} }  =  \frac{ \frac{\partial^{2} }{ \partial \beta^{2}}  \left [ \left ( \frac{ \hbar k_{B} c^5 }{ 4 \pi G } \right )   \beta^{2}  \right ] }{ k_{B} }  =   \frac{ \hbar  c^5 }{ 2 \pi G }   ,
\label{variance_SBH}
\end{equation}
where $\mathcal{E}$ and $S$ in Eq.\ (\ref{variance}) have been replaced by $\mathcal{E}_{H}$ and $S_{H}$, respectively.
The replacement is assumed to be valid.
Substituting the Planck energy $E_{P} = \sqrt{ \hbar c^{5} / G  }$ given by Eq.\ (\ref{eq:Mp-Ep}) into Eq.\ (\ref{variance_SBH}) yields
\begin{equation}
 \sigma_{\mathcal{E}_{H}}^{2}   
                                      =   \frac{ \hbar  c^5 }{ 2 \pi G }   =  \frac{1}{2 \pi }    E_{P}^{2}    .
\label{variance_SBH_Ep}
\end{equation}
This equation indicates that the thermodynamic energy fluctuations are a universal constant corresponding to the Planck energy.
That is, the thermodynamic fluctuations do not depend on the Hubble parameter $H$, although de Sitter universes depend on $H$.
The universality implies that the thermodynamic fluctuations of the energy on the horizon are characterized by the Planck energy.
(Quantum fluctuations characterized by the Planck length were studied in Ref.\ \cite{Ford1997}.)

We examine the case of the event horizon of a Schwarzschild black hole in Appendix\ \ref{Case of an event horizon of a Schwarzschild black hole}, to compare with the Hubble horizon.
As shown in Eq.\ (\ref{variance_S_bh_Ep}), thermodynamic fluctuations of the energy on the event horizon are also a universal constant corresponding to the Planck energy.
Accordingly, the thermodynamic fluctuations on the two horizons are equivalent to each other, excepting the numerical coefficients.

Next, we examine relative fluctuations of the energy on the Hubble horizon.
For this, we use the average energy $ \left < \mathcal{E}_{H}  \right >$ given by Eq.\ (\ref{E_ST2}), based on the holographic equipartition law.
From Eqs.\ (\ref{E_ST2}) and (\ref{variance_SBH}), the square of the relative fluctuations is written as 
\begin{align}
 \frac{  \sigma_{\mathcal{E}_{H}}^{2}    }{\left <   \mathcal{E}_{H}  \right >^{2}  }  
  &=   \frac{ \frac{ \hbar  c^5 }{ 2 \pi G } }{ \left ( \frac{ c^{5} }{ G }  \left ( \frac{1}{H} \right )  \right )^{2}  }   = \frac{    k_{B}    }{ 2  \left ( \frac{ \pi k_{B} c^5 }{ \hbar G } \right )  \frac{1}{H^2} }   \notag \\
  &=   \frac{ k_{B} }{ 2 } \frac{ 1 }{  S_{H} }  = \frac{ 2 }{ N_{H} }   ,
\label{variance_Enegy}
\end{align}
where $S_{H}  = \left ( \frac{ \pi k_{B} c^5 }{ \hbar G } \right )  \frac{1}{H^2}$ given by Eq.\ (\ref{eq:SBH3}) and $ N_{H} = 4 S_{H} / k_{B} $ given by Eq.\ (\ref{N_sur}) have been used.
From the above equation, we can confirm that the square of the relative fluctuations corresponds to the inverse of $N_{H}$. 
An equivalent result has been discussed in previous works, see, e.g., Ref.\ \cite{CosmoFluc0}.
(Equation\ (\ref{variance_Enegy}) is consistent with Eq.\ (\ref{variance_Enegy_bh}) for the event horizon of a Schwarzschild black hole.)
In addition, using Eq.\ (\ref{variance_SBH_Ep}), $ \sigma_{\mathcal{E}_{H}}^{2}   /  \left <   \mathcal{E}_{H}  \right >^{2} $ can be written as
\begin{align}
 \frac{ \sigma_{\mathcal{E}_{H}}^{2}  }{\left <   \mathcal{E}_{H}  \right >^{2}  }   
 &= \frac{  \frac{1}{2 \pi }   E_{P}^{2} }{  \left <   \mathcal{E}_{H}  \right >^{2}  }                                                       
   = \frac{  \frac{1}{2 \pi }   (E_{P}/c^{2})^{2} }{  ( \left <   \mathcal{E}_{H}  \right >/c^{2})^{2}  }     = \frac{1}{2 \pi } \frac{ M_{P}^{2}  }{ M_{H}^{2} }           ,
\label{variance_Enegy_3}
\end{align}
or equivalently, from Eq.\ (\ref{variance_Enegy}), we obtain 
\begin{align}
 \frac{ \sigma_{\mathcal{E}_{H}}^{2}  }{\left <   \mathcal{E}_{H}  \right >^{2}  }   
 &=   \frac{ \frac{ \hbar  c^5 }{ 2 \pi G } }{ \left ( \frac{ c^{5} }{ G }  \left ( \frac{1}{H} \right )  \right )^{2}  }                              
  = \frac{1}{2 \pi }  \left (  \frac{ \hbar G }{ c^5 }  \right ) H^{2}   = \frac{1}{2 \pi } \frac{ t_{P}^{2}  }{ t_{H}^{2} }                                                     \notag \\
 &= \frac{1}{2 \pi }  \left (  \frac{ \hbar G }{ c^3 }  \right ) \left ( \frac{H}{c} \right )^{2}   =  \frac{1}{2 \pi } \frac{ L_{P}^{2}  }{ r_{H}^{2} }    .
\label{variance_Enegy_4}
\end{align}
Here $L_{P}$, $ t_{P}$, and $M_{P}$ represent the Planck length, time, and mass, respectively, which are summarized in Eqs.\ (\ref{eq:Lp-tp}) and (\ref{eq:Mp-Ep}).
Also, $r_{H}$ is the Hubble radius given by $c/H$ from Eq.\ (\ref{eq:rH}), $ t_{H}$ is the Hubble time defined by $1/H$, and $M_{H}$ is the Hubble mass defined by $ \left < \mathcal{E}_{H}  \right >/c^{2}$.
From Eqs.\ (\ref{variance_Enegy_3}) and (\ref{variance_Enegy_4}), the relative fluctuations of the energy on the Hubble horizon are summarized as
\begin{align}
 \frac{ \sigma_{\mathcal{E}_{H}} }{\left <   \mathcal{E}_{H}  \right >  }  \approx  \frac{ E_{P} }{ E_{H} }  =  \frac{ M_{P} }{ M_{H}  }   =  \frac{ t_{P} }{ t_{H} }  =  \frac{ L_{P} }{ r_{H}  }   ,
\label{relative_2}
\end{align}
where $\left < \mathcal{E}_{H} \right > = E_{H}$ has been used for the Hubble energy.
In this way, the relative fluctuations of energy can be characterized by the ratio of the Planck energy (mass, time, and length) to the Hubble energy (mass, time, and radius).
Temperature is not included here, but is discussed in the next paragraph.

Applying $H =  2 \pi  k_{B}T_{H} / \hbar$ given by Eq. (\ref{eq:T_H1}) to Eq.\ (\ref{variance_Enegy_4}) yields
\begin{align}
 \frac{ \sigma_{\mathcal{E}_{H}}^{2}  }{\left <   \mathcal{E}_{H}  \right >^{2}  }     
 &= \frac{1}{2 \pi }  \left (  \frac{ \hbar G }{ c^5 }  \right ) H^{2}   
  =  \frac{1}{2 \pi }  \left (  \frac{ \hbar G }{ c^5 }  \right ) \left ( \frac{   2 \pi  k_{B} T_{H} } { \hbar} \right )^{2}                                              \notag \\
 &= 8 \pi \left (  \frac{ G }{ \hbar  c^5 }  \right ) \left ( \frac{1}{2} k_{B} T_{H} \right )^{2}   =  8 \pi  \left (  \frac{ \frac{1}{2} k_{B} T_{H}  }{  E_{P} }  \right )^{2}     ,
\label{variance_Enegy_5}
\end{align}
where Eq.\ (\ref{eq:Mp-Ep}) has been used for the Planck energy $E_{P} = \sqrt{ \hbar c^{5} / G  }$.
Therefore, the relative fluctuations of energy are written as
\begin{align}
 \frac{ \sigma_{\mathcal{E}_{H}}  }{\left <   \mathcal{E}_{H}  \right >  }     &=  2 \sqrt{ 2\pi}  \left (  \frac{ \frac{1}{2} k_{B} T_{H}  }{  E_{P} }  \right )   \approx   \frac{ \frac{1}{2} k_{B} T_{H}  }{  E_{P} }    .
\label{relative_3}
\end{align}
Here $\frac{1}{2} k_{B} T_{H}$ corresponds to the energy for 1-DOF, assuming the holographic equipartition law of energy.
The relative fluctuations are characterized by the ratio of the 1-DOF energy to the Planck energy.
Using the Planck temperature $T_{P} =E_{P}/k_{B}$, Eq.\ (\ref{relative_3}) is given as
\begin{align}
 \frac{ \sigma_{\mathcal{E}_{H}}  }{\left <   \mathcal{E}_{H}  \right >  }     &= 2 \sqrt{ 2\pi}  \left (  \frac{ \frac{1}{2} k_{B} T_{H}  }{  E_{P} }  \right )   =  \sqrt{ 2\pi}  \left (  \frac{ T_{H}  }{  T_{P} }  \right )  \approx  \frac{ T_{H}  }{  T_{P} }      .
\label{relative_4}
\end{align}
The relative fluctuations correspond to $T_{H}/T_{P}$.

We note that the relative fluctuations for the Hubble horizon examined here are consistent with those for the event horizon of a Schwarzschild black hole.
For example, Eqs.\ (\ref{relative_2}), (\ref{relative_3}), and (\ref{relative_4}) are consistent with Eqs.\ (\ref{relative_bh}), (\ref{relative_Enegy_bh}), and (\ref{relative_Enegy_bh_2}), respectively.

Finally, we discuss current values of the relative fluctuations on the Hubble horizon.
At the present time, the value of $\sigma_{\mathcal{E}_{H}}^{2} / \left <   \mathcal{E}_{H}  \right >^{2}$ is approximately $10^{-123}$, which is calculated from substituting $S_{H}/ k_{B} = 2.6 \times 10^{122}$ \cite{Egan1} into Eq.\ (\ref{variance_Enegy}).
($N_{H}$ at the present time is approximately $10^{123}$. The number of bits of the Universe has been discussed in Refs.\ \cite{Barrow1999,Lloyd2002}.)
Consequently, we obtain $  \sigma_{\mathcal{E}_{H}}   /  \left <   \mathcal{E}_{H}  \right > \approx 10^{-62}$ within the range $\rho_{\Lambda,obs} / \rho_{vac,th}$, which is approximately from $10^{-60}$ to $10^{-123}$ \cite{Weinberg1989etc}.
Here, $\rho_{\Lambda,obs}$ and $\rho_{vac,th}$ represent the mass density for the observed value of $\Lambda$ and the mass density for the theoretical value of the vacuum energy, respectively.
The relation should be approximately written as 
\begin{align}
 \frac{ \sigma_{\mathcal{E}_{H}}  }{\left <   \mathcal{E}_{H}  \right >  }     &\approx    10^{-62}     \sim \frac{\rho_{\Lambda,obs} }{ \rho_{vac,th} }   ,
\label{relative_L_vac}
\end{align}
where the symbol `$\sim$' represents an approximation based on a rough estimation that includes a larger uncertainty than `$\approx$'.
As discussed in Sec.\ \ref{Energy on the horizon}, the order of $\rho_{H}$ calculated from $\left <  \mathcal{E}_{H} \right >$ is consistent with the order of $\rho_{\Lambda,obs}$.
From Eq.\ (\ref{rhoH_rhoLobs}), the relation is written as
\begin{equation}
 \rho_{H} \approx \rho_{\Lambda,obs}  .
\label{rhoH_rhoLobs_2}
\end{equation}
Therefore, the relative fluctuations examined here may lead to a discussion on the discrepancy between $\rho_{\Lambda,obs}$ and $\rho_{vac,th}$.
(Note that $\sigma_{\mathcal{E}_{H}}$ itself should be much smaller than the observed energy for $\Lambda$.)
For example, the relative energy fluctuations given by Eq.\ (\ref{relative_4}) can be written as 
\begin{align}
 \frac{ \sigma_{\mathcal{E}_{H}}  }{\left <   \mathcal{E}_{H}  \right >  }     &\approx \frac{ T_{H} }{ T_{P} }  =   \frac{ N_{H} \times \frac{1}{2} k_{B} T_{H}  }{ N_{H} \times \frac{1}{2} k_{B} T_{P}  }    =   \frac{ E_{H} }{ E_{HP} }  ,
\label{relative_E_Emax}
\end{align}
where Eq.\ (\ref{E_equip}) has been used for $ E_{H}$.
Also, $E_{HP}$ represents the energy that can be `maximally' stored on the Hubble horizon, which is assumed to be given by 
\begin{equation}
 E_{HP} = N_{H} \times \frac{1}{2} k_{B} T_{P}    ,
\label{eq:E_HP}
\end{equation}
where $T_{P}$ is the Planck temperature.
In Eq.\ (\ref{eq:E_HP}), $\frac{1}{2} k_{B} T_{P}$ is assumed to be the 1-DOF energy which can be maximally stored on a Planckian area, in order to apply the holographic equipartition law.
That is, an extended holographic equipartition law is assumed, where the maximum 1-DOF energy corresponds to $E_{P}/2$.
These assumptions have not been established and should be beyond the limits of validity.
However, as a viable scenario, we accept the assumptions here.
Using Eqs.\ (\ref{relative_L_vac}) and (\ref{relative_E_Emax}), the order of the relative fluctuations should be approximately written as 
\begin{align}
 \frac{ \sigma_{\mathcal{E}_{H}}  }{\left <   \mathcal{E}_{H}  \right >  }  &\approx \frac{ E_{H} }{ E_{HP} } = \frac{ \rho_{H} }{ \rho_{HP} }  \sim \frac{\rho_{\Lambda,obs} }{ \rho_{vac,th} }   ,
\label{relative_E_Emax_H_Lobs}
\end{align}
where $\rho_{H}$ and $\rho_{HP}$ represent mass density calculated from $E_{H} /(c^{2} V_{H})$ and $E_{HP} /(c^{2} V_{H})$, respectively.
Substituting Eq.\ (\ref{rhoH_rhoLobs_2}) into Eq.\ (\ref{relative_E_Emax_H_Lobs}) and arranging the resultant equation yields
\begin{equation}
  \rho_{HP} \sim \rho_{vac,th}          .
\label{rho_HP-rho_vac}
\end{equation}
The above calculation is based on the rough estimation and the unestablished assumptions.
However, the maximum energy that can be stored on the Hubble horizon may behave as if it were a kind of effective vacuum-like energy in the extended holographic equipartition law.

In fact, it has been reported that the vacuum fluctuations of energy density are consistent with the observed $\Lambda$ \cite{Padma2005}, as described in Sec.\ \ref{Introduction}.
Also, assuming a thermal density matrix, vacuum energy fluctuations on a spacetime geometry are characterized by an entanglement entropy \cite{Verlinde2020}, as if they are thermodynamic fluctuations.
Accordingly, the energy stored on the Hubble horizon, its thermodynamic fluctuations, and the maximum energy examined in the present paper may be related to the vacuum energy fluctuations reported in the previous works.
We may be able to study this relation by extending the idea of entanglement entropy discussed in AdS/CFT \cite{Verlinde2020,RyuTakayanagi2006,Zurek20202021}.
This task is left for future research.

\section{Conclusions}
\label{Conclusions}

We examined the energy stored on a horizon and its thermodynamic fluctuations through the holographic equipartition law.
Stable fluctuations around thermal equilibrium states were assumed, to study the thermodynamic energy fluctuations.

First, we confirmed that a flat FRW universe should approach a de Sitter universe at least in the last stage, using a cosmological model close to $\Lambda$CDM models.
Then, based on the holographic equipartition law, we calculated the energy density of the Hubble volume from the energy stored on the Hubble horizon of a de Sitter universe.
The energy density is constant and the order of the energy density is consistent with the order of that for $\Lambda$ by observations, namely $\rho_{H} \approx \rho_{\Lambda,obs}$.
As a viable scenario, the energy stored on the horizon may be related to a kind of effective dark energy, through the holographic equipartition law.

Second, we examined thermodynamic fluctuations of the energy on the horizon, assuming stable fluctuations around thermal equilibrium states.
A standard formulation of the fluctuations for a canonical ensemble was applied to the Hubble horizon of a de Sitter universe.
The thermodynamic fluctuations of the energy on the Hubble horizon are found to be a universal constant corresponding to the Planck energy.
That is, the thermodynamic fluctuations depend only on the Planck energy, regardless of the Hubble parameter.
By applying the holographic equipartition law, the relative energy fluctuations can be characterized by the ratio of the Planck energy (mass, time, and length) to the Hubble energy (mass, time, and radius).
Also, the relative fluctuations correspond to the ratio of the 1-DOF energy to the Planck energy.
The thermodynamic fluctuations and relative fluctuations are consistent with those for the event horizon of a Schwarzschild black hole.

Finally, we discussed current values of the relative fluctuations of the energy on the Hubble horizon.
At the present time, the relative fluctuations are approximately $10^{-62}$.
The order of the fluctuations can be approximately written as $\sigma_{\mathcal{E}_{H}} / \left <  \mathcal{E}_{H}  \right > \approx \rho_{H} / \rho_{HP} \sim \rho_{\Lambda,obs} / \rho_{vac,th}$  by assuming an (unestablished) extended holographic equipartition law.
In addition, $\rho_{HP} \sim \rho_{vac,th}$ should be expected.
The relative fluctuations and the energy that can be `maximally' stored on the Hubble horizon may be related to a kind of effective vacuum-like energy in the extended holographic equipartition law.

This study has revealed fundamental properties of thermodynamic fluctuations of the energy on the horizon, assuming thermal equilibrium.
The present results should provide new insights into alternative dark energy and may lead to a discussion on the cosmological constant problem.
Of course, we cannot exclude other contributions such as quantum field theory.
Detailed studies are needed and these are left for future research.

\appendix

\section{Case of an event horizon of a Schwarzschild black hole}
\label{Case of an event horizon of a Schwarzschild black hole}

Thermodynamic fluctuations of the energy on an event horizon of a Schwarzschild black hole have been studied from various viewpoints \cite{Fluctuations1,Fluctuations2,Canonial0,Das2002,More2005,Pourhassan2017,Stability1,AbreuNeto2020,OthersBlackholes,Nojiri2003}.
In this appendix, we examine the thermodynamic fluctuations on an event horizon, to compare with the case of the Hubble horizon.
We assume stable fluctuations around thermal equilibrium states, as examined in Sec.\ \ref{Fluctuations of energy on the Hubble horizon} and previous works \cite{Das2002,Nojiri2003,More2005,Pourhassan2017}.

In Appendix\ \ref{Thermodynamics on an event horizon}, we review the thermodynamics on the event horizon of a black hole and discuss several thermodynamic relations.
In Appendix\ \ref{Fluctuations of energy on the event horizon}, we examine thermodynamic fluctuations of the energy on an event horizon.

\subsection{Thermodynamics on an event horizon of a Schwarzschild black hole}
\label{Thermodynamics on an event horizon}

We briefly review the thermodynamics on the event horizon of a Schwarzschild black hole \cite{Hawking2}.
Using the Bekenstein--Hawking entropy, the black hole entropy is written as 
\begin{equation}
 S_{bh}  = \frac{ k_{B} c^3 }{  \hbar G }  \frac{A_{bh}}{4}   ,
\label{eq:Sbh}
\end{equation}
where $A_{bh}$ is the surface area of the sphere with the event horizon $r_{bh}$ and the subscript `$bh$' represents the black hole.
The black hole radius $r_{bh}$ is given by
\begin{equation}
     r_{bh} = \frac{2GM_{bh}}{c^{2}}   ,
\label{eq:Rbh}
\end{equation}
where $M_{bh}$ is the mass of the black hole.
Substituting $A_{bh}=4 \pi r_{bh}^2 $ into Eq.\ (\ref{eq:Sbh}) and applying Eq.\ (\ref{eq:Rbh}) yields
\begin{align}
S_{bh} &= \frac{ k_{B} c^3 }{  \hbar G }  \frac{A_{bh}}{4} = \frac{ k_{B} c^3 }{  \hbar G }   \frac{4 \pi r_{bh}^2    }{4}      
            = \frac{  \pi  k_{B} c^3 }{  \hbar G }  \left ( \frac{2GM_{bh}}{c^{2}}  \right )^2       \notag \\
         &=  \left ( \frac{ 4 \pi k_{B} G }{ \hbar c } \right )  M_{bh}^2  .
\label{eq:Sbh2}      
\end{align}
The temperature $T_{bh}$ on the event horizon of the black hole is given by \cite{Hawking2}
\begin{equation}
 T_{bh} = \frac{ \hbar c^{3}}{   8 \pi  k_{B} G M_{bh} }   .
\label{eq:T_bh}
\end{equation}
This equation indicates that $T_{bh}$ depends on $M_{bh}$.
When $M_{bh}$ is constant, $T_{bh}$ is constant.
Consequently, the event horizon is considered to be in thermal equilibrium, namely $T_{bh}=T_{0}$.
In this case, thermodynamic fluctuations of the energy on the horizon can be calculated from Eq.\ (\ref{variance}), assuming stable fluctuations around thermal equilibrium states.

Before discussing fluctuations, we examine several thermodynamic relations using the average energy, to compare with the case of the Hubble horizon.
Multiplying Eq.\ (\ref{eq:Sbh2}) by Eq.\ (\ref{eq:T_bh}) yields 
\begin{align}
 S_{bh} T_{bh} &= \left ( \frac{ 4 \pi k_{B} G }{ \hbar c } \right )  M_{bh}^2  \times  \frac{ \hbar c^{3}}{   8 \pi  k_{B} GM_{bh} }      \notag \\
             &= \frac{1}{2} M_{bh} c^{2}      = \frac{1}{2}  \left < \mathcal{E}_{bh} \right >    ,
\label{eq:S_bhT_bh}
\end{align}
where the average energy $\left < \mathcal{E}_{bh}  \right >$ of the black hole is assumed to be given by
\begin{equation}
 \left < \mathcal{E}_{bh} \right > = M_{bh} c^{2} .
\label{eq:EMc2_bh}
\end{equation}
Therefore, Eq.\ (\ref{eq:S_bhT_bh}) is written as
\begin{align}
\left < \mathcal{E}_{bh}  \right > = 2 S_{bh} T_{bh} . 
\label{eq:E_S_bhT_bh}
\end{align}
This relation is consistent with $ \left < \mathcal{E}_{H} \right > =2 S_{H} T_{H} $ given by Eq.\ (\ref{E_ST}) for the Hubble horizon.
Next, let us examine whether or not a formulation of the holographic equipartition law $\left < \mathcal{E}_{bh} \right > =  N_{bh}  \times \frac{1}{2} k_{B} T_{bh} $ is satisfied.
To this end, we assume that the number $N_{bh}$ of DOF on the event horizon is given by Eq.\ (\ref{N_sur}).
Using Eq.\ (\ref{N_sur}) and replacing $N_{H}$ and $S_{H}$ by $N_{bh}$ and $S_{bh}$, respectively, we obtain
\begin{equation}
  N_{bh} =  \frac{4 S_{bh} }{k_{B}}   ,     \quad \textrm{or equivalently} \quad       S_{bh} =  \frac{ k_{B} N_{bh}}{4}       .
\label{N_sur_bh}
\end{equation}
Substituting Eq.\ (\ref{N_sur_bh}) into Eq.\ (\ref{eq:E_S_bhT_bh}) yields 
\begin{equation}
\left < \mathcal{E}_{bh} \right > =  N_{bh}  \times \frac{1}{2} k_{B} T_{bh}     .
\label{E_equip_bh}
\end{equation}
This confirms the formulation of the holographic equipartition law, as expected.
The holographic equipartition law has been discussed in the works of Padmanabhan \cite{Padma2010} and Verlinde \cite{Verlinde1}.
In addition, let us confirm the thermodynamic relation $\frac{  d \left < \mathcal{E}_{bh} \right > }{ T_{bh} }          =   d S_{bh}$.
Taking $ d \left < \mathcal{E}_{bh} \right > = dM_{bh} c^{2}$ obtained from Eq.\ (\ref{eq:EMc2_bh}) and dividing $ d \left < \mathcal{E}_{bh} \right >$ by Eq.\ (\ref{eq:T_bh}) yields
\begin{equation}
\frac{  d \left < \mathcal{E}_{bh} \right > }{ T_{bh} }          =  \left ( \frac{ 8 \pi k_{B} G }{ \hbar c } \right )  M_{bh}   dM_{bh}        =   d S_{bh}     .
\end{equation}
Here $d S_{bh} =  \left ( \frac{ 8 \pi k_{B} G }{ \hbar c } \right ) M_{bh}   dM_{bh}$ is obtained from Eq.\ (\ref{eq:Sbh2}).
In this way, we can confirm the two thermodynamic relations and the formulation of the holographic equipartition law, as for the case of the Hubble horizon.

\subsection{Thermodynamic fluctuations of the energy on the event horizon of a Schwarzschild black hole}
\label{Fluctuations of energy on the event horizon}

We now study thermodynamic fluctuations of the energy on the event horizon of a Schwarzschild black hole.
From Eq.\ (\ref{variance}), the variance $\sigma_{\mathcal{E}_{bh}}^{2}$ of the energy on an event horizon is written as
\begin{equation}
 \sigma_{\mathcal{E}_{bh}}^{2} \equiv \left <  \left ( \delta \mathcal{E}_{bh} \right )^{2} \right >  =\frac{ S_{bh}^{\prime \prime} (\beta) }{ k_{B} } ,
\label{variance_bh}
\end{equation}
where $\mathcal{E}$ and $S$ in Eq.\ (\ref{variance}) have been replaced by $\mathcal{E}_{bh}$ and $S_{bh}$, respectively.
To apply this equation, $S_{bh}$ given by Eq.\ (\ref{eq:Sbh2}) is formulated as a function of $\beta$.
Using Eq.\ (\ref{eq:T_bh}), the inverse temperature $\beta$ can be written as
\begin{equation}
\beta  = \frac{1}{k_{B} T} = \frac{1}{k_{B} T_{bh}}= \left ( \frac{ 8 \pi G  }{  \hbar c^{3}  } \right ) M_{bh}  ,
\label{eq_beta_bh}
\end{equation}
where $T=T_{0}=T_{bh}$ is assumed.
From Eq.\ (\ref{eq_beta_bh}),  $M_{bh}$ is given by
\begin{equation}
M_{bh} = \left ( \frac{ \hbar c^{3}  }{ 8 \pi G  } \right ) \beta  .
\label{eq_M_beta_bh}
\end{equation}
Substituting Eq.\ (\ref{eq_M_beta_bh}) into Eq.\ (\ref{eq:Sbh2}) yields 
\begin{align}
S_{bh}   &=  \left ( \frac{ 4 \pi k_{B} G }{ \hbar c } \right )  \left ( \frac{ \hbar c^{3}  }{ 8 \pi G  } \right )^{2}  \beta^{2}   
              =   k_{B} \left ( \frac{ \hbar c^{5}  }{ 16 \pi G  } \right )  \beta^{2} .
\label{eq:Sbh_beta}      
\end{align}
In addition, substituting Eq.\ (\ref{eq:Sbh_beta}) into Eq.\ (\ref{variance_bh}) yields
\begin{equation}
 \sigma_{\mathcal{E}_{bh}}^{2}  
                             = \frac{ \frac{\partial^{2}  }{ \partial \beta^{2} } \left [     k_{B} \left ( \frac{ \hbar c^{5}  }{ 16 \pi G  } \right )  \beta^{2}  \right ] }{ k_{B} } =   \frac{ \hbar  c^5 }{ 8 \pi G }  
                             =   \frac{   E_{P}^{2} }{ 8 \pi  }   ,
\label{variance_S_bh_Ep}
\end{equation}
where we have used the Planck energy, $E_{P} = \sqrt{ \hbar c^{5} / G  }$, given by Eq.\ (\ref{eq:Mp-Ep}).
The thermodynamic fluctuations of the energy on the event horizon are a universal constant corresponding to the Planck energy, as for the case of the Hubble horizon.
That is, the thermodynamic fluctuations do not depend on the $M_{bh}$ that characterizes the Schwarzschild black hole.
From Eqs.\ (\ref{variance_SBH_Ep}) and (\ref{variance_S_bh_Ep}), the energy fluctuations are given by $ \sigma_{\mathcal{E}_{H}} =    E_{P} / \sqrt{2 \pi } $ and $ \sigma_{\mathcal{E}_{bh}} =    E_{P} / (2\sqrt{2 \pi }) $, respectively.
Accordingly, the thermodynamic fluctuations of the energy on the two horizons are equivalent to each other, excepting the numerical coefficients.
This universality may prompt a discussion on dark energy and the microscopic structures of spacetime, through an entanglement entropy related to black hole thermodynamics \cite{Verlinde2020,RyuTakayanagi2006,Zurek20202021}.

Finally, we examine the relative fluctuations of the energy on an event horizon.
To this end, $ \left < \mathcal{E}_{bh} \right > $ given by Eq.\ (\ref{eq:EMc2_bh}) is considered to be the energy on the event horizon.
From Eqs.\ (\ref{eq:EMc2_bh}) and (\ref{variance_S_bh_Ep}), the square of the relative fluctuations is written as 
\begin{align}
 \frac{  \sigma_{\mathcal{E}_{bh}}^{2}    }{\left <   \mathcal{E}_{bh}  \right >^{2}  }  
           &=   \frac{ \frac{ \hbar  c^5 }{ 8 \pi G } }{  ( M_{bh} c^{2} )^{2}  }   =  \frac{ k_{B} }{ 2 }  \frac{ 1 }{ \left ( \frac{4 \pi k_{B} G }{ \hbar c } \right )  M_{bh}^2  }      \notag \\
           & =   \frac{ k_{B} }{ 2 } \frac{ 1 }{ S_{bh} }  = \frac{ 2 }{ N_{bh} }   ,
\label{variance_Enegy_bh}
\end{align}
where we have used $S_{bh}  = \left ( \frac{ 4 \pi k_{B} G }{ \hbar c } \right )  M_{bh}^2$ given by Eq.\ (\ref{eq:Sbh2}) and $ N_{bh} = 4 S_{bh} / k_{B} $ given by Eq.\ (\ref{N_sur_bh}).
The square of the relative fluctuations corresponds to the inverse of the number of DOF.
This result agrees with Eq.\ (\ref{variance_Enegy}) for the Hubble horizon.
In addition, using Eqs.\ (\ref{eq:EMc2_bh}) and (\ref{variance_S_bh_Ep}) and performing several operations, the relative fluctuations can be summarized as 
\begin{align}
 \frac{ \sigma_{\mathcal{E}_{bh}} }{\left <   \mathcal{E}_{bh}  \right >  }    =\frac{1}{\sqrt{8 \pi}} \frac{ E_{P} }{ E_{bh} }  =  \frac{1}{\sqrt{8 \pi}} \frac{ M_{P} }{ M_{bh}  }   = \frac{1}{\sqrt{2 \pi}}   \frac{ t_{P} }{ t_{bh} }  = \frac{1}{\sqrt{2 \pi}}  \frac{ L_{P} }{ r_{bh}  }   , 
\label{relative_bh}
\end{align}
where $E_{bh}$ represents $\left < \mathcal{E}_{bh}  \right >$ and $t_{bh}$ is a crossing time defined by $r_{bh}/c$.
Also, $M_{P}$, $t_{P}$, and $L_{P}$ represent the Planck mass, time, and length, respectively.
Equation\ (\ref{relative_bh}) indicates that the relative fluctuations of energy can be characterized by the ratio of the Planck energy (mass, time, and length) to the black-hole energy (mass, time, and radius).
Accordingly, Eq.\ (\ref{relative_bh}) is consistent with Eq.\ (\ref{relative_2}) for the Hubble horizon when the numerical coefficients included in Eq.\ (\ref{relative_bh}) are neglected.
To discuss the temperature, substituting Eq.\ (\ref{eq_M_beta_bh}) into $ \left < \mathcal{E}_{bh} \right > = M_{bh} c^{2}$ and applying $E_{P} = \sqrt{ \hbar c^{5} / G  }$ yields 
\begin{align}
 \left < \mathcal{E}_{bh} \right > &= M_{bh} c^{2} = \left ( \frac{ \hbar c^{3}  }{ 8 \pi G  } \right ) \beta c^{2}   = \left ( \frac{ \hbar c^{5}  }{ 8 \pi G  } \right ) \beta     \notag \\
                                 &=  \left ( \frac{ E_{P}^{2} }{ 8 \pi   } \right ) \beta .
\label{eq:E_beta_Epl}
\end{align}
Dividing the square root of Eq.\ (\ref{variance_S_bh_Ep}) by Eq.\ (\ref{eq:E_beta_Epl}) and applying $\beta = 1/(k_{B} T_{bh})$ yields
\begin{align}
 \frac{ \sigma_{\mathcal{E}_{bh}}   }{\left <  \mathcal{E}_{bh}  \right >  }    & = \frac{ \sqrt{ \frac{   E_{P}^{2} }{ 8 \pi  } }  }{ \left ( \frac{ E_{P}^{2} }{ 8 \pi   } \right ) \beta }  = 4 \sqrt{2 \pi}  \left ( \frac{  \frac{1}{2} k_{B} T_{bh}  }{  E_{P} }  \right )  \approx  \frac{  \frac{1}{2} k_{B} T_{bh}  }{  E_{P} }  ,
\label{relative_Enegy_bh}
\end{align}
where $\frac{1}{2} k_{B} T_{bh}$ corresponds to the $1$-DOF energy.
The relative energy fluctuations are characterized by the ratio of the 1-DOF energy to the Planck energy.
Applying the Planck temperature $T_{P} =E_{P}/k_{B}$ to Eq.\ (\ref{relative_Enegy_bh}) yields
\begin{align}
 \frac{ \sigma_{\mathcal{E}_{bh}}   }{\left <  \mathcal{E}_{bh}  \right >  }    & = 4 \sqrt{2 \pi} \left ( \frac{ \frac{1}{2} k_{B} T_{bh}  }{  E_{P} }  \right ) = 2 \sqrt{2 \pi} \left ( \frac{  T_{bh}   }{  T_{P} }  \right )  \approx \frac{  T_{bh}   }{  T_{P} }  .
\label{relative_Enegy_bh_2}
\end{align}
Equations\ (\ref{relative_Enegy_bh}) and (\ref{relative_Enegy_bh_2}) are consistent with Eqs.\ (\ref{relative_3}) and (\ref{relative_4}), respectively.
(Note that the numerical coefficients for the former two equations are slightly different from those for the latter two.)
Using Eq.\ (\ref{relative_Enegy_bh_2}), the relative fluctuations are written as 
\begin{align}
 \frac{ \sigma_{\mathcal{E}_{bh}}  }{\left <   \mathcal{E}_{bh}  \right >  }     &\approx \frac{ T_{bh} }{ T_{P} }   =   \frac{ N_{bh} \times \frac{1}{2} k_{B} T_{bh}  }{ N_{bh} \times   \frac{1}{2} k_{B} T_{P} }    =   \frac{ E_{bh}  }{ E_{bhP} }  ,
\label{relative_vac_bh}
\end{align}
where $E_{bh}$ represents $N_{bh} \times \frac{1}{2} k_{B} T_{bh}$ given by Eq.\ (\ref{E_equip_bh}).
Also, $E_{bhP}$ represents the energy that can be `maximally' stored on the event horizon, which is assumed to be given by $E_{bhP}= N_{bh} \times \frac{1}{2} k_{B} T_{P}$.
Based on this assumption, Eq.\ (\ref{relative_vac_bh}) is consistent with Eq.\ (\ref{relative_E_Emax}) for the Hubble horizon.
We note that the assumption has not been established, as mentioned in Sec.\ \ref{Fluctuations of energy on the Hubble horizon}.

The thermodynamic relations, thermodynamic fluctuations, and relative fluctuations for the Hubble horizon of a de Sitter universe are confirmed to be consistent with those for the event horizon of a Schwarzschild black hole, using the framework considered in the present study.
This consistency should provide a deeper understanding of horizon thermodynamics.


\begin{thebibliography}{99}

\bibitem{PERL1998_Riess1998} S. Perlmutter \textit{et al.},  Nature (London) \textbf{391}, 51 (1998); A. G. Riess \textit{et al.}, Astron. J. \textbf{116}, 1009 (1998). 
\bibitem{Planck2015} P. A. R. Ade \textit{et al.}, Astron. Astrophys. \textbf{594}, A13 (2016).     
\bibitem{Planck2018} N. Aghanim \textit{et al.}, Astron. Astrophys. \textbf{641}, A6 (2020).        
\bibitem{Hubble2017} O. Farooq, F. R. Madiyar, S. Crandall, B. Ratra, Astrophys. J. \textbf{835}, 26 (2017).     

\bibitem{Bamba1} K. Bamba, S. Capozziello, S. Nojiri, S. D. Odintsov, Astrophys. Space Sci. \textbf{342}, 155 (2012).



\bibitem{Freese-Mimoso_2015}
K. Freese, F. C. Adams, J. A. Frieman, E. Mottola, Nucl. Phys. \textbf{B287}, 797 (1987); 
J. M. Overduin, F. I. Cooperstock, Phys. Rev. D \textbf{58}, 043506 (1998).
\bibitem{Sola_2009-2018} 
S. Basilakos, M. Plionis, J. Sol\`{a}, Phys. Rev. D \textbf{80}, 083511 (2009);
S. Basilakos, A. Paliathanasis, J. D. Barrow, G. Papagiannopoulos, Eur. Phys. J. C \textbf{78}, 684 (2018).
%
\bibitem{Nojiri2006}  
S. Nojiri, S. D. Odintsov, Phys. Lett. B \textbf{639}, 144 (2006); 
Q. Wang, Z. Zhu, W. G. Unruh, Phys. Rev. D \textbf{95}, 103504 (2017).
%
%
\bibitem{Richarte2013a} 
L. P. Chimento, M. G. Richarte, Phys. Rev. D \textbf{84}, 123507 (2011);
L. P. Chimento, M. G. Richarte, Iv\'{a}n E. S\'{a}nchez Garc\'{i}a, Phys. Rev. D \textbf{88}, 087301 (2013).
%
\bibitem{Sola_2015L14}
J. Sol\`{a}, A. G\'{o}mez-Valent, J. C. P\'{e}rez, Astrophys. J. \textbf{811}, L14 (2015).
\bibitem{Valent2015} 
A. G\'{o}mez-Valent, J. Sol\`{a}, S. Basilakos, J. Cosmol. Astropart. Phys. 01 (2015) 004.
%
\bibitem{Sola2019} 
M. Rezaei, M. Malekjani, J. Sol\`{a} Peracaula, Phys. Rev. D \textbf{100}, 023539 (2019).
%
\bibitem{Sola_2017-2020} 
J. Sol\`{a}, A. G\'{o}mez-Valent, J. C. P\'{e}rez, Phys. Lett. B \textbf{774}, 317 (2017);
S. Basilakos, N. E. Mavromatos, J. Sol\`{a} Peracaula, Phys. Rev. D \textbf{101}, 045001 (2020).




%
\bibitem{Weinberg0} S. Weinberg, \textit{Gravitation and Cosmology} (John Wiley \& Sons, New York, 1972).
\bibitem{Murphy1}   G. L. Murphy, Phys. Rev. D \textbf{8}, 4231 (1973).
\bibitem{Barrow11-Zimdahl1}  J. D. Barrow, Phys. Lett. B  \textbf{180}, 335 (1986);
J. A. S. Lima, R. Portugal, I. Waga,  Phys. Rev. D \textbf{37}, 2755 (1988).
%
\bibitem{Brevik1-Nojiri1} 
I. Brevik, S. D. Odintsov,  Phys. Rev. D \textbf{65}, 067302 (2002);
S. Nojiri, S. D. Odintsov, Phys. Rev. D \textbf{72}, 023003 (2005);
I. Brevik, E. Elizalde, S. Nojiri, S. D. Odintsov, Phys. Rev. D \textbf{84}, 103508 (2011).
%
\bibitem{Barrow21} B. Li,  J. D. Barrow, Phys. Rev. D \textbf{79}, 103521 (2009).
%
\bibitem{Meng3} X. Dou, X.-H. Meng, Advances in Astronomy \textbf{2011}, 829340 (2011).
%
\bibitem{Avelino2etc2020} 
A. Avelino, U. Nucamendi, J. Cosmol. Astropart. Phys. 04 (2009) 006; 
A. Sasidharan, N. D. J. Mohan, M. V. John, T. K. Mathew, Eur. Phys. J. C \textbf{78}, 628 (2018); 
W. Yang, S. Pan, E. DiValentino, A. Paliathanasis, J. Lu, Phys. Rev. D \textbf{100} 103518 (2019).




\bibitem{Easson}  D. A. Easson, P. H. Frampton, G. F. Smoot, Phys. Lett. B \textbf{696}, 273 (2011).
\bibitem{Cai} Y. F. Cai, E. N. Saridakis,  Phys. Lett. B \textbf{697}, 280 (2011).
\bibitem{Basilakos1} S. Basilakos, D. Polarski, J. Sol\`{a}, Phys. Rev. D \textbf{86}, 043010 (2012).
\bibitem{Basilakos2014-Gohar}
S. Basilakos, J. Sol\`{a}, Phys. Rev. D \textbf{90}, 023008 (2014); 
M. P. D\c{a}browski, H. Gohar, Phys. Lett. B \textbf{748}, 428 (2015); 
R. C. Nunes, E. M. Barboza Jr., E. M. C. Abreu, J. A. Neto, J. Cosmol. Astropart. Phys. 08 (2016) 051.
%
\bibitem{Sheykhi1}
A. Sheykhi, Phys. Rev. D \textbf{81}, 104011 (2010).
\bibitem{Sadjadi1} 
H. M. Sadjadi, M. Jamil, Europhys. Lett. \textbf{92}, 69001 (2010); 
S. Mitra, S. Saha, S. Chakraborty, Mod. Phys. Lett. A \textbf{30}, 1550058 (2015).
%
\bibitem{Sheykhi2} A. Sheykhi, S. H. Hendi, Phys. Rev. D \textbf{84}, 044023 (2011). 
\bibitem{Karami2011-2016} 
K. Karami, A. Abdolmaleki, Z. Safari, S. Ghaffari, J. High Energy Phys. 08 (2011) 150.


\bibitem{Koma4}   N. Komatsu, S. Kimura, Phys. Rev. D \textbf{87}, 043531 (2013); N. Komatsu, JPS Conf. Proc. \textbf{1}, 013112 (2014).
\bibitem{Koma5}   N. Komatsu, S. Kimura, Phys. Rev. D \textbf{88}, 083534 (2013).
\bibitem{Koma678}  N. Komatsu, S. Kimura, Phys. Rev. D \textbf{89}, 123501 (2014); Phys. Rev. D \textbf{90}, 123516 (2014); Phys. Rev. D \textbf{92}, 043507 (2015).
\bibitem{Koma9}  N. Komatsu, S. Kimura, Phys. Rev. D \textbf{93}, 043530 (2016).



\bibitem{Hooft-Bousso}
G. 't Hooft, Conf. Proc. C \textbf{930308}, 284 (1993) [arXiv:gr-qc/9310026]; L. Susskind, J. Math. Phys. \textbf{36}, 6377 (1995); R. Bousso, Rev. Mod. Phys. \textbf{74}, 825 (2002).


%
\bibitem{Bekenstein1}  
J. D. Bekenstein, Phys. Rev. D \textbf{7}, 2333 (1973); Phys. Rev. D \textbf{9}, 3292 (1974);  Phys. Rev. D \textbf{12}, 3077 (1975).
\bibitem{Hawking1}  
S. W. Hawking, Phys. Rev. Lett. \textbf{26}, 1344 (1971); Commun. Math. Phys. \textbf{43}, 199 (1975); Phys. Rev. D \textbf{13}, 191 (1976).
\bibitem{Hawking2}  
S. W. Hawking, Nature (London) \textbf{248}, 30 (1974). 


\bibitem{Jacob1995} T. Jacobson, Phys. Rev. Lett. \textbf{75}, 1260 (1995).
\bibitem{Padma2010}    
T. Padmanabhan, Mod. Phys. Lett. A \textbf{25}, 1129 (2010).
\bibitem{Verlinde1} 
E. Verlinde, J. High Energy Phys. 04 (2011) 029.
%
\bibitem{HDE}  
M. Li, Phys. Lett. B \textbf{603}, 1 (2004);
A. Sayahian Jahromi, S. A. Moosavi, H. Moradpour, J. P. Morais Gra\c{c}a, I. P. Lobo, I. G. Salako, A. Jawad, Phys. Lett. B \textbf{780}, 21 (2018);
S. Nojiri, S. D. Odintsov, T. Paul, Symmetry \textbf{13}, 928 (2021).
%
%
%
\bibitem{Padmanabhan2004}  
T. Padmanabhan, Classical Quantum Gravity \textbf{21}, 4485 (2004).
%
\bibitem{ShuGong2011}  
Fu-Wen Shu, Y. Gong, Int. J. Mod. Phys. D \textbf{20}, 553 (2011).



\bibitem{Padma2012AB}  T. Padmanabhan, arXiv:1206.4916 [hep-th]; Res. Astron. Astrophys. \textbf{12}, 891 (2012).


\bibitem{Cai2012-Tu2013}                                                                               
R. G. Cai, J. High Energy Phys. 1211 (2012) 016.
\bibitem{Tu2013-2015}  
S. Chakraborty, T. Padmanabhan, Phys. Rev. D \textbf{92}, 104011 (2015); 
H. Moradpour, Int. J. Theor. Phys. \textbf{55}, 4176 (2016). 

\bibitem{Sheykhia2018} 
A. Sheykhi, Phys. Lett. B \textbf{785}, 118 (2018);
Phys. Rev. D \textbf{103}, 123503 (2021).
\bibitem{Neto2018a} E. M. C. Abreu, J. A. Neto, A. C. R. Mendes, A. Bonilla, Europhys. Lett. \textbf{121}, 45002 (2018).


\bibitem{Koma10}  N. Komatsu, Eur. Phys. J. C \textbf{77}, 229 (2017).
\bibitem{Koma11}  N. Komatsu, Phys. Rev. D \textbf{96}, 103507 (2017).
\bibitem{Koma12}  N. Komatsu, Phys. Rev. D \textbf{99}, 043523 (2019).

\bibitem{Koma14}  N. Komatsu, Phys. Rev. D \textbf{100}, 123545 (2019).
\bibitem{Koma15}  N. Komatsu, Phys. Rev. D \textbf{102}, 063512 (2020).
\bibitem{Koma16}  N. Komatsu, Phys. Rev. D \textbf{103}, 023534 (2021).




\bibitem{Easther1}  R. Easther, D. Lowe, Phys. Rev. Lett. \textbf{82}, 4967 (1999).
\bibitem{Barrow3}  J. D. Barrow, New Astronomy \textbf{4}, 333 (1999).
\bibitem{Davies11_Davis0100} 
 T. M. Davis, P. C. W. Davies, C. H. Lineweaver, Classical Quantum Gravity \textbf{20}, 2753 (2003). 
\bibitem{Gong00_01}    B. Wang, Y. Gong, E. Abdalla, Phys. Rev. D \textbf{74}, 083520 (2006).
\bibitem{Egan1} C. A. Egan, C. H. Lineweaver, Astrophys. J. \textbf{710}, 1825 (2010).

\bibitem{Pavon2013Mimoso2013}
D. Pav\'{o}n, N. Radicella, Gen. Relativ. Gravit. \textbf{45}, 63 (2013);
J. P. Mimoso, D. Pav\'{o}n, Phys. Rev. D \textbf{87}, 047302 (2013).
%
\bibitem{Krishna20172019} P. B. Krishna, T. K. Mathew, Phys. Rev. D \textbf{96}, 063513 (2017);
Phys. Rev. D \textbf{99}, 023535 (2019).
%
\bibitem{Bamba2018}  K. Bamba, A. Jawad, S. Rafique, H. Moradpour, Eur. Phys. J. C \textbf{78}, 986 (2018).
\bibitem{Pavon2019}
M. Gonzalez-Espinoza, D. Pav\'{o}n, Mon. Not. R. Astron. Soc. \textbf{484}, 2924 (2019).



\bibitem{deSitter}  
L. Dyson, M. Kleban, L. Susskind, J. High Energy Phys. 10 (2002) 011;
A. Albrecht, J. Phys. Conf. Ser. \textbf{174}, 012006 (2009);
S. M. Carroll, A. Chatwin-Davies, Phys. Rev. D \textbf{97}, 046012 (2018).


\bibitem{Saridakis2019}
S. Pan, W. Yang, C. Singha, E. N. Saridakis, Phys. Rev. D \textbf{100}, 083539 (2019).
%
\bibitem{Saridakis2021}
E. N. Saridakis, S. Basilakos, Eur. Phys. J. C \textbf{81} 644 (2021).





\bibitem{Fluctuations1}      
G. W. Gibbons, M. J. Perry, Phys. Rev. Lett. \textbf{36}, 985 (1976);
S. W. Hawking, D. N. Page, Commun. Math. Phys. \textbf{87}, 577 (1983);
D. Pav\'{o}n, J. M. Rub\'{i}, Phys. Rev. D \textbf{37}, 2052 (1988).
\bibitem{Fluctuations2} 
A. Casher, F. Englert, N. Itzhaki, S. Massar, R. Parentani, Nucl. Phys. B \textbf{484}, 419 (1997);
A. Chamblin, R. Emparan, C. V. Johnson, R. C. Myers, Phys. Rev. D. \textbf{60}, 104026 (1999);
G. Gour, Phys. Rev. D \textbf{61}, 021501(R) (1999);
G. Gour, A. J. M. Medved, Classical Quantum Gravity \textbf{20}, 3307 (2003).
\bibitem{Canonial0}       
A. Chatterjee, P. Majumdar, Phys. Rev. Lett. \textbf{92}, 141301 (2004).
%
\bibitem{OthersBlackholes}       
M. Faizal, A. Ashour, M. Alcheikh, L. Alasfar, S. Alsaleh, A. Mahroussah, Eur. Phys. J. C \textbf{77}, 608 (2017);
W. S. Chung, H. Hassanabadi, Phys. Lett. B \textbf{793}, 451 (2019);
A. Ghosh, S. Mukherji, C. Bhamidipati, arXiv:2104.12720 [hep-th].


\bibitem{Das2002}  
S. Das, P. Majumdar, R. K. Bhaduri, Classical Quantum Gravity \textbf{19}, 2355 (2002).
%
\bibitem{Nojiri2003}
S. Nojiri, S. D. Odintsov, S. Ogushi, Int. J. Mod. Phys. A \textbf{18}, 3395 (2003).
%
\bibitem{More2005}  
S. S. More, Classical Quantum Gravity \textbf{22}, 4129 (2005).
\bibitem{Pourhassan2017} 
B. Pourhassan, K. Kokabi, S. Rangyan, Gen. Relativ. Gravit. \textbf{49}, 144 (2017).
%
%
\bibitem{Stability1}       
M-S. Ma, R. Zhao, Phys. Lett. B \textbf{751}, 278 (2015);
T. S. Bir\'{o}, V. G. Czinner, H. Iguchi, P. V\'{a}n, Phys. Lett. B \textbf{782}, 228 (2018);
K. Mejrhit, S-E. Ennadifi, Phys. Lett. B \textbf{794}, 45 (2019).


\bibitem{AbreuNeto2020}  
E. M. C. Abreu, J. A. Neto, Phys. Lett. B \textbf{810}, 135805 (2020);
Eur. Phys. J. C \textbf{80}, 776 (2020);  
E. M. C. Abreu, J. A. Neto, E. M. Barboza Jr., Europhys. Lett. \textbf{130}, 40005 (2020).



\bibitem{Mimoso2018}
J. P. Mimoso, D. Pav\'{o}n, Phys. Rev. D \textbf{97}, 103537 (2018).


\bibitem{CosmoFluc0}
M. Artymowski, J. Mielczarek, Eur. Phys. J. C \textbf{79}, 632 (2019).



%
\bibitem{Peebles_1993} P. J. E. Peebles, \textit{Principles of Physical Cosmology} (Princeton University Press, Princeton, New Jersey, 1993).
%
\bibitem{Fluc_others1}
C. Bennett \textit{et al.}, Astrophys. J. \textbf{436}, 423 (1994); 
S. Bharadwaj, S. S. Ali, Mon. Not. R. Astron. Soc. \textbf{352}, 142 (2004).
%
\bibitem{Fluc_others2}
J. Magueijo, L. Pogosian, Phys. Rev. D \textbf{67}, 043518 (2003);
J. Magueijo, P. Singh, Phys. Rev. D \textbf{76}, 023510 (2007);
J. Magueijo, L. Smolin, C. R. Contaldi, Classical Quantum Gravity \textbf{24}, 3691 (2007).
%
\bibitem{Fluc_others3}
W-J. Li, Y. Ling, J-P. Wu, X-M. Kuang, Phys. Lett. B \textbf{687}, 1 (2010);
T. Biswas, R. Brandenberger, T. Koivisto, A. Mazumdar, Phys. Rev. D \textbf{88}, 023517 (2013);
Z. Haba, Eur. Phys. J. C \textbf{78}, 596 (2018).





\bibitem{CosmoFluc_Quantum}
A. Vilenkin, Nucl. Phys. B \textbf{226}, 527 (1983); 
Y. Nambu, Phys. Rev. D \textbf{78}, 044023 (2008);
M. Maggiore, Phys. Rev. D \textbf{83}, 063514 (2011).
%
%
\bibitem{Ford1997}
L. H. Ford, N. F. Svaiter, Phys. Rev. D \textbf{56}, 2226 (1997).
%
%
\bibitem{Padma2005}  T. Padmanabhan, Classical Quantum Gravity \textbf{22}, L107 (2005).

\bibitem{Elizalde2005}
E. Elizalde, S. Nojiri, S. D. Odintsov, P. Wang, Phys. Rev. D \textbf{71}, 103504 (2005).

\bibitem{Verlinde2020}
E. Verlinde, K. M. Zurek, J. High Energy Phys. \textbf{04} 209 (2020).



\bibitem{Weinberg1989etc} 
S. Weinberg, Rev. Mod. Phys. \textbf{61}, 1 (1989);  
V. Sahni, A. A. Starobinsky, Int. J. Mod. Phys. D \textbf{9}, 373 (2000); 
S. M. Carroll, Living Rev. Relativity \textbf{4}, 1 (2001); 
T. Padmanabhan, Phys. Rep. \textbf{380}, 235 (2003);
R. Bousso, Gen. Relativ. Gravit. \textbf{40}, 607 (2008);
J. D. Barrow, D. J. Shaw, Phys. Rev. Lett. \textbf{106}, 101302, (2011); 
%
J. Martin, C. R. Physique \textbf{13}, 566 (2012);
N. Bao, R. Bousso, S. Jordan, B. Lackey, Phys. Rev. D \textbf{96}, 103512 (2017);
N. G. Sanchez, Phys. Rev. D \textbf{104}, 123517 (2021).





%
\bibitem{Das2008Radicella2010} S. Das, S. Shankaranarayanan, S. Sur, Phys. Rev. D \textbf{77}, 064013 (2008);
N. Radicella, D. Pav\'{o}n, Phys. Lett. B \textbf{691}, 121 (2010).
%
%
\bibitem{MeissnerGhosh} 
K. A. Meissner, Classical Quantum Gravity \textbf{21}, 5245, (2004); 
A. Ghosh, P. Mitra, Phys. Rev. D \textbf{71}, 027502, (2005).
%
\bibitem{Tsallis2012}  C. Tsallis, L. J. L. Cirto, Eur. Phys. J. C \textbf{73}, 2487 (2013). 
%
\bibitem{Czinner1}       T. S. Bir\'{o}, V. G. Czinner, Phys. Lett. B \textbf{726}, 861 (2013).
\bibitem{Czinner2}    V. G. Czinner, H. Iguchi, Phys. Lett. B \textbf{752}, 306 (2016);
Eur. Phys. J. C \textbf{77}, 892 (2017).
%
\bibitem{Barrow2020} J. D. Barrow, Phys. Lett. B \textbf{808}, 135643 (2020).


\bibitem{Nojiri2021}
S. Nojiri, S. D. Odintsov, V. Faraoni, Phys. Rev. D \textbf{104}, 084030 (2021).



\bibitem{GibbonsHawking1977} 
G. W. Gibbons, S. W. Hawking, Phys. Rev. D \textbf{15}, 2738 (1977).


\bibitem{Evans1993etc}
D. J. Evans, E. G. D. Cohen, G. P. Morriss, Phys. Rev. Lett. \textbf{71}, 2401 (1993); 
G. Gallavotti, E. G. D. Cohen, Phys. Rev. Lett. \textbf{74}, 2694 (1995).


\bibitem{Barrow1999} J. D. Barrow, New Astronomy \textbf{4}, 333 (1999).
\bibitem{Lloyd2002}  S. Lloyd, Phys. Rev. Lett. \textbf{88}, 237901 (2002).



\bibitem{RyuTakayanagi2006}
S. Ryu, T. Takayanagi, Phys. Rev. Lett. \textbf{96} 181602 (2006);
J. High Energy Phys. \textbf{08}, 045 (2006).
%
\bibitem{Zurek20202021}
K. M. Zurek, Phys. Lett. B \textbf{826}, 136910 (2022); T. Banks, K. M. Zurek, arXiv:2108.04806 [hep-th].







\end{thebibliography}
\end{document}